# Effect of transient change in strain rate on plastic flow behavior of low carbon steel

A. Ray, P. Barat*, P. Mukherjee, A. Sarkar and S.K. Bandyopadhyay
Variable Energy Cyclotron Centre,
1/AF, Bidhan Nagar, Kolkata-700064, INDIA

**Abstract**

Plastic flow behavior of low carbon steel has been studied at room temperature during tensile deformation by varying the initial strain rate of $3.3 \times 10^{-4}$ $s^{-1}$ to the final strain rate ranging from $1.33\times10^{-3}$ $s^{-1}$ to $2.0\times10^{-3}$ $s^{-1}$ at a fixed engineering strain of 12%. *Haasen plot* revealed that the mobile dislocation density remained almost invariant at the juncture where there was a sudden increase in stress with the change in strain rate and the plastic flow was solely dependent on the velocity of mobile dislocations. In that critical regime, the variation of stress with time was fitted with a *Boltzman* type Sigmoid function. The increase in stress was found to increase with final strain rate and the time elapsed to attain these stress values showed a decreasing trend. Both of these parameters saturated asymptotically at higher final strain rate.

* Corresponding author. Tel: +91-33-2337-1230, Fax: +91-33-2334-6871, Email: pbarat@veccal.ernet.in

# 1. Introduction

The plasticity of metals and alloys is a consequence of the motion of the line defects or dislocations, under the action of an applied stress [1-3]. Their motion is obstructed by point defects, impurities, electrons, phonons, other dislocations and the Peierls barrier [4] that originates from the crystal structure itself. The stress-strain curve of the materials in the plastic regime is thus governed by the dynamics of dislocations. The plastic deformation of solids is often described by the equations based on the variables of the deformation state of the form $F(\sigma, \varepsilon, \dot{\varepsilon}, T) = 0$, where $\sigma$ is the mean true stress including the internal stress, $\varepsilon$ is the true strain, $\dot{\varepsilon}$ is the true strain rate, and $T$ is the temperature. Furthermore, one can express the true strain rate, $\dot{\varepsilon}$, as a function of $\sigma$ for fixed $\varepsilon$ and T.

Several experimental techniques are adopted in order to study the plastic deformation. Some of these techniques include strain rate change test, stress change test and stress rate change test [5,6] within a short interval of time. The aim of these deformation transient experiments is to study the deformation kinetics and the associated microstructural changes. In the present work we have carried out strain rate change test on low carbon steel and shown the effect of sudden change of strain rate on the plastic flow behavior primarily governed by the velocity of mobile dislocations.

# 2. Experiments and results

The composition of the low carbon steel used in the present study was (in wt. %): 0.15 C, 0.33 Mn, 0.04 P, 0.05 S, 0.15 Si and rest Iron. The gauge length of the cylindrical tensile specimens was 25 mm. The uniaxial tensile tests were performed at room temperature with the help of a servo controlled INSTRON (model 4482) machine. The initial crosshead speed was 0.5mm/min (corresponding to engineering strain rate of value $3.3\times10^{-4}s^{-1}$) for all the specimens up to an engineering strain of 12%. The cross head speed was then changed to different values, *i.e.*

2,4,5,6,8,10,12,15,18,20,25 and 30 mm/min. The tensile load and the displacement were measured using a computer-controlled interface at a sampling rate of 20 points/sec.

Fig. 1 shows a segment of a typical true stress *vs.* true strain curve in the plastic regime showing a sudden increase in true stress at 11.33% true strain (corresponding to 12% engineering strain) due to sudden increase in the strain rate. The yield stress values for the specimens were found to lie within 460-480 MPa. Prior to strain hardening, extensive serration were noticed in all the specimens. This oscillation of stress is associated with the heterogeneous deformation behavior of the material, mainly due to the barrier-controlled dynamics of dislocations [7]. The present study was carried out in the plastic regime where the deformation is nearly homogeneous.

## 3. The strain rate sensitivity analysis

It is known that, an increase in strain rate increases the tensile strength of metals [8]. The sudden change in the strain rate during tensile deformation could be a very useful approach to understand the dynamics of dislocations. The strain rate sensitivity, *m*, which is defined as $m = \ln(\sigma_2 / \sigma_1) / \ln(\dot{\varepsilon}_2 / \dot{\varepsilon}_1)$, where subscripts '1' and '2' stand for the initial and final values respectively, was calculated and the values were found to be ~0.01 for all the samples. A *Haasen plot* [9], $\Delta\sigma / (T\ \Delta \ln\dot{\varepsilon})$ *vs.* $\sigma$-$\sigma_Y$, which is commonly used to study the nature of dislocation interaction in monotonically deformed materials [10], obtained from the true stress-true strain data for all the samples is shown in Fig. 2. Here, $\Delta\sigma$ is the change of stress due to a change in the strain rate $\dot{\varepsilon}$; $\sigma$-$\sigma_Y$ is the effective flow stress, $\sigma_Y$ is the yield stress and *T* is the temperature.

Generally metals at room temperature, with few regularly spaced dislocations exhibit linear *Haasen plot* passing through the origin. Its slope is proportional to the inverse of the *operational* activation area [11] and is expected to vary as the number of mobile dislocation changes with time [12]. However, since the strain rate sensitivity, *m*, is proportional to the slope of the *Haasen plot*, the apparent constancy in the parameter *m* suggested that the mobile dislocation density did not change appreciably during the transient changes of strain rate [10].

Again it is reported that [13], the dislocation density, $\rho$, in metals evolves with strain, ε, by the following the relation:

$$\rho = \rho_0 + M\varepsilon^{\beta} \tag{1}$$

where $\rho_0$ is the initial dislocation density, $M$ is the multiplication factor that does not depend either on the strain rate or temperature and $\beta$ is the strain exponent, assumed to be unity in most cases. This relation justifies that the total dislocation density after a sudden increase in strain rate remains almost unaltered at a fixed strain value. So it is straightforward to argue that, during the transient change in strain rate, the possibility of generation of significant amount of mobile dislocations could be ruled out.

According to *Orowan equation* [11], the plastic strain rate, $\dot{\varepsilon}$, is expressed in terms of mobile dislocation density, $\rho_m$ and velocity, $v$ as

$$\dot{\varepsilon} = \rho_m b v \tag{2}$$

where $b$ is the magnitude of Burger's vector. As $\rho_m$ is constant, a sudden increase in the strain rate has caused an increase in the dislocation velocity only.

The true stress, σ, at the true strain value of 11.33%, increased for all the cases when the strain rate was changed. The variation of σ with time was plotted (Fig. 3) and found to fit well with the *Boltzman* type Sigmoid function of the form,

$$\sigma(t) = \frac{\sigma_i - \sigma_f}{\left(1 + e^{\frac{t-t_0}{\Delta t}}\right)} + \sigma_f \tag{3}$$

Here, $\sigma_i$ and $\sigma_f$ are the stress values for $t<t_0$ and $t>t_0$ respectively, $t_0$ being the time at which $\sigma(t_0) = (\sigma_i + \sigma_f)/2$ and $\Delta t$ is related to the width of the Sigmoid. The best fitted values of the parameters $\Delta t$ and ($\sigma_f$-$\sigma_i$) for different final strain rate ($\dot{\varepsilon}_2$) are shown in Fig.4. From Fig.4, it is interesting to note that the increment in stress, ($\sigma_f$-$\sigma_i$) increased and Δt decreased with $\dot{\varepsilon}_2$. Both of them showed asymptotic type of saturation at higher $\dot{\varepsilon}_2$.

The strain rate dependence of ($\sigma_f$-$\sigma_i$) can be explained as follows. Using equation 2 and assuming the dislocation velocity, $v = A\sigma^{m'}$ ($A$ is constant) [14] we obtained

$$\ln \dot{\varepsilon} = \ln(A\rho_m\ b) + m' \ln \sigma \qquad (4)$$

As $\rho_m$ remained constant during transient change in strain rate, it can be shown that,

$$\sigma_f - \sigma_i = <\sigma_i> \left(\frac{\dot{\varepsilon}_2}{\dot{\varepsilon}_1}\right)^{1/m'} - <\sigma_i> \qquad (5)$$

where $<\sigma_i>$ is the average initial true stress. Using $<\sigma_i>$=561.16 Mpa for all the tested samples and fitting eq. (5) with the data shown Fig. 4, we got $m^{/}$ = 97.08 which lead to [14] $m = \frac{1}{m'}$ = 0.0103. This provided an excellent match with the average strain rate sensitivity of 0.01 obtained from the experiments. The Δt followed a power law behavior with final strain rate of the form $\Delta t = \Delta t_0 + C(\dot{\varepsilon}_2)^{\alpha}$, where $\Delta t_0 = 12.2\times10^{-3}$ s, C = 0.0004 and α = -0.7. This phenomenological power law dependence was only possible as the velocity of dislocations saturates with increasing stress field.

## 4. Conclusion

The stress-strain measurements were carried out in the low-carbon steel specimen at room temperature following a sudden strain rate change in its plastic deformation regime. The strain rate sensitivities were calculated for different final strain rates. It was found from the analysis of Haasen plot that the transient change in the strain rate in plastic regime associated a change in the average dislocation velocity instead of the change in the dislocation density.

**FIGURE CAPTIONS:**

Fig. 1. A segment of typical true stress vs. true strain curve in the plastic regime of the low carbon steel specimen for initial crosshead speed 0.5mm/min and a final cross head speed of 10.0 mm/min.

Fig. 2. *Haasen plot* for the specimens at room temperature within the error of ±5%.

Fig. 3. The experimental data of true stress with time (open circles) at the juncture of increase in final strain rate fitted with a Sigmoid function. The solid line is the fitted curve.

Fig. 4. Variation of Δt and ($\sigma_f$-$\sigma_i$) with final strain rate values.

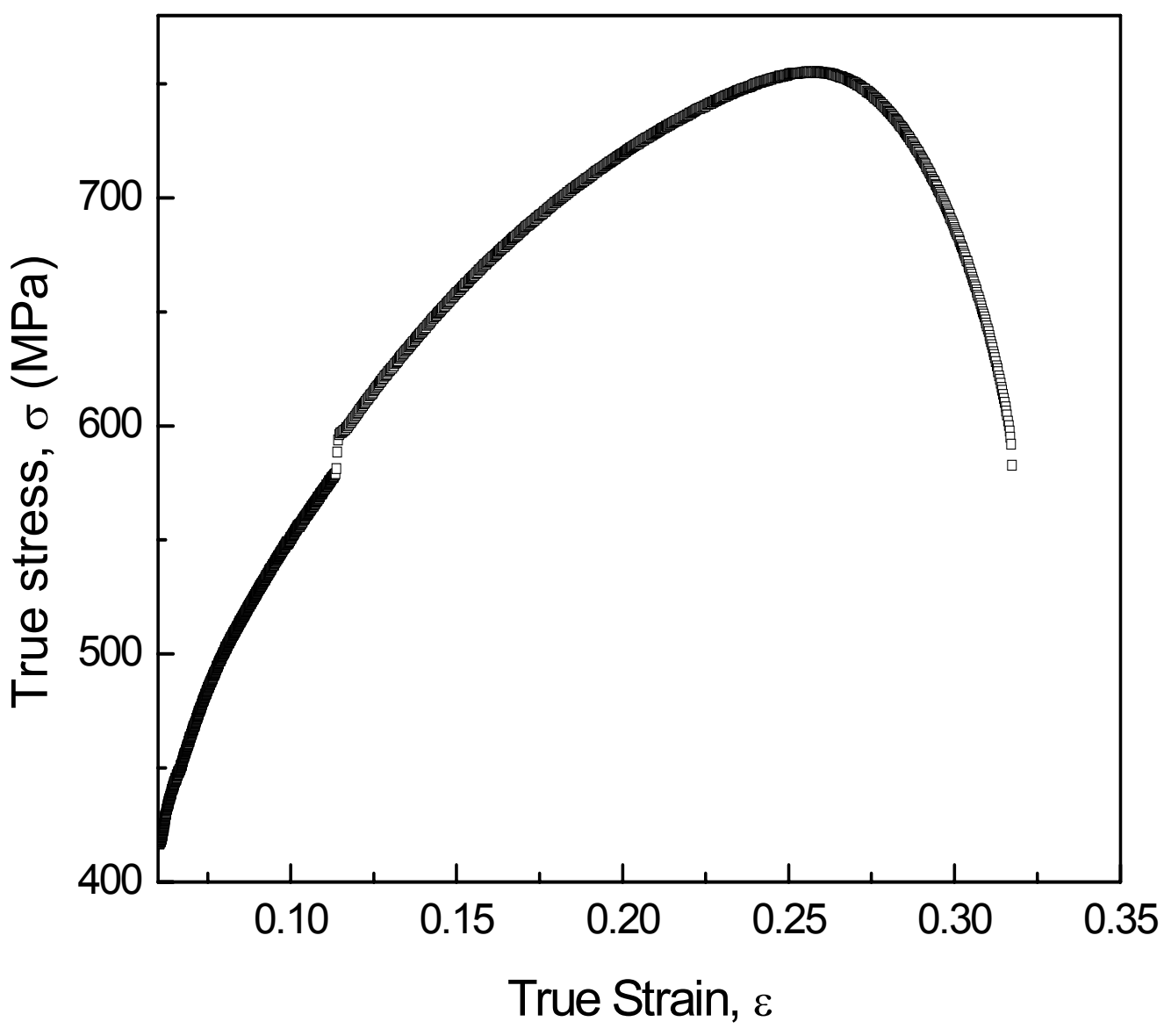

**Fig. 1**

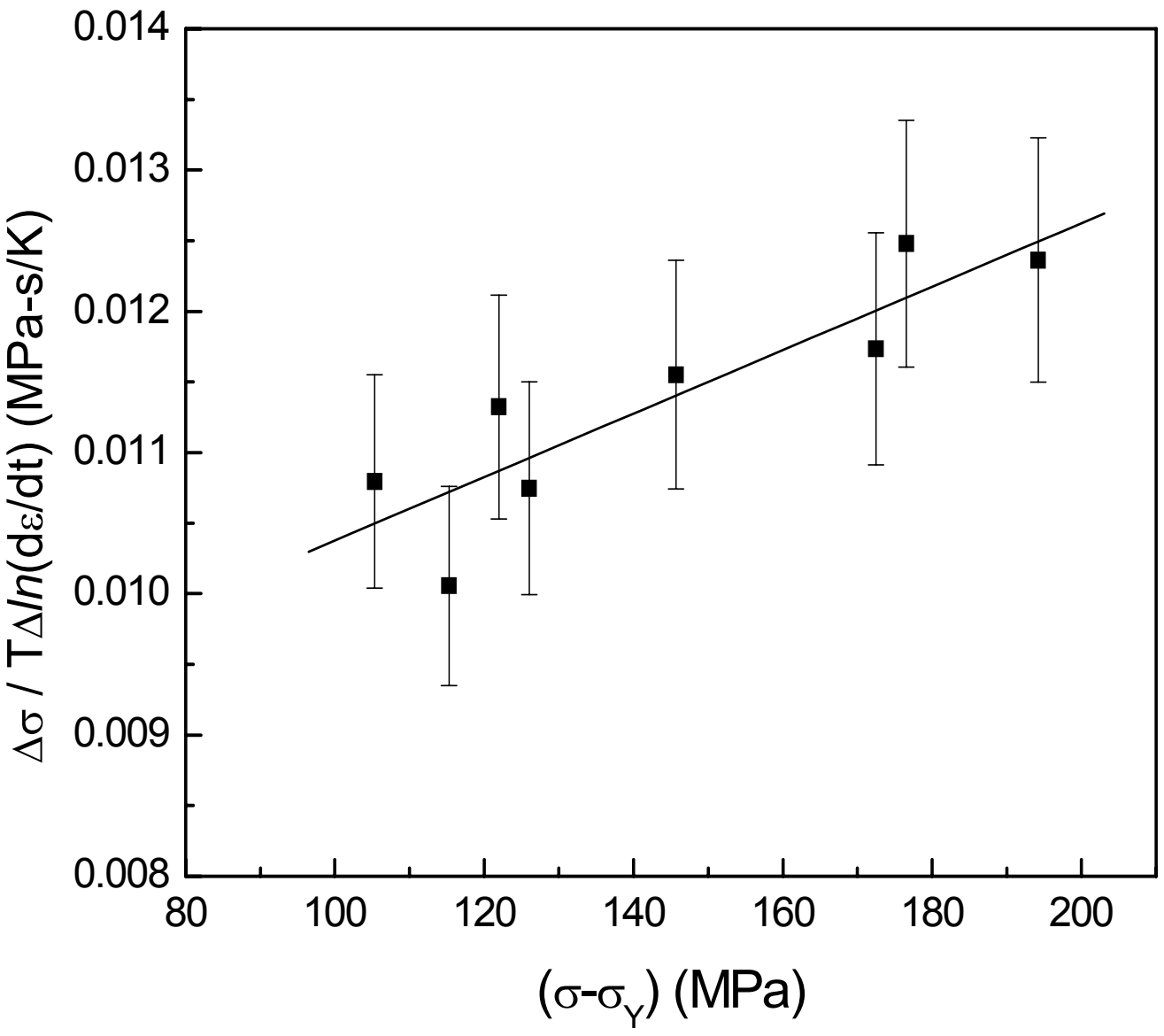

**Fig. 2**

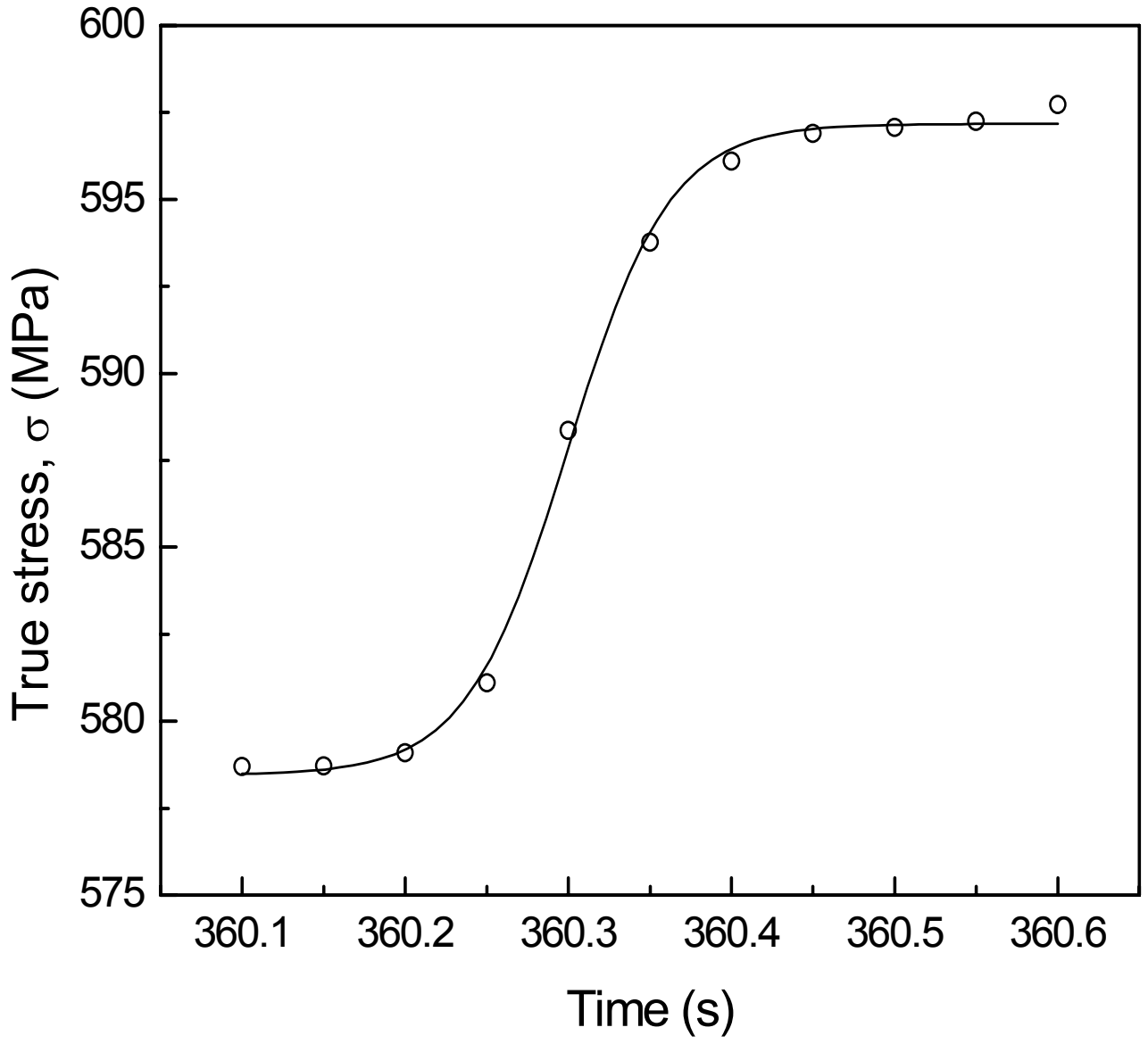

**Fig. 3**

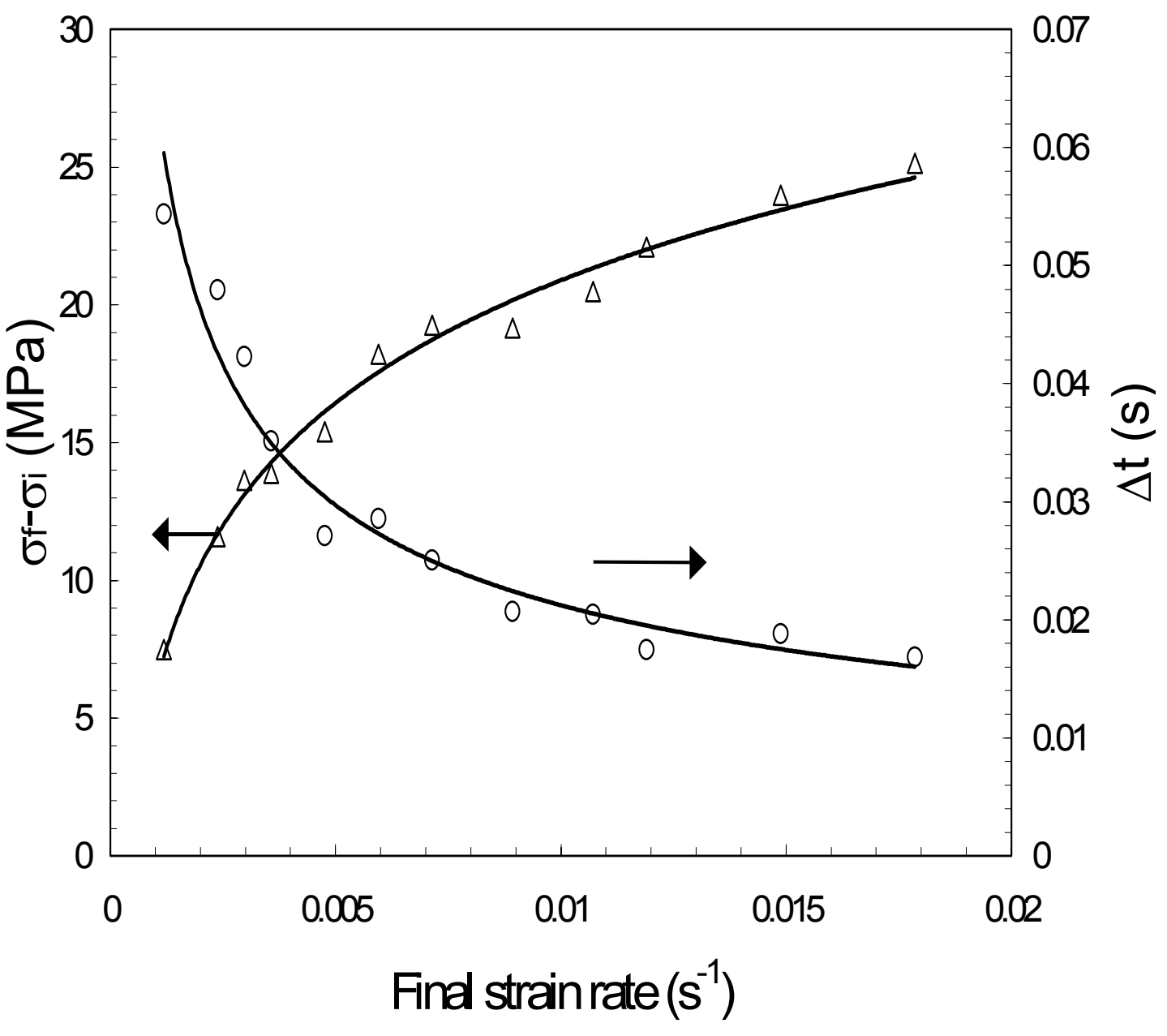

**Fig. 4**